\documentclass{elsart}

\usepackage{amssymb}

\newcommand{\mgb}{MgB$_{2}$}

\begin{document}

\begin{frontmatter}



\title{Improved superconducting properties of \mgb}


\author{N. A. Frederick, Shi Li, M. B. 
Maple}
\address{Department of Physics and Institute for Pure and Applied
Physical Sciences,\\
University of California, San Diego, La Jolla, CA 92093}

\author{V. F. Nesterenko, S. S. Indrakanti}
\address{Department of Mechanical and Aerospace Engineering,\\
University of California, San Diego, La Jolla, CA 92093}

\date\today

\begin{abstract}
We present electrical transport, magnetization, and specific heat
measurements on bulk \mgb{} samples ($T_{c} = 38.5$ K) synthesized
under $200$ MPa pressure using a process based on hot isostatic pressing 
with cooling under pressure. 
The samples are fully dense and display excellent superconducting 
properties, including a narrow superconducting transition width 
($\Delta{}T_{c} = 0.75$ K), a high upper critical field $H_{c2}$ 
($H_{c2}(0) \sim 155$ kOe), and a critical current density 
$J_{c}$ that is the largest yet measured for bulk samples of \mgb{}
($J_{c}(0) \sim 1.4$ MA/cm$^{2}$).  Specific heat 
measurements yielded a jump $\Delta{}C$ at $T_{c}$ of $92$ mJ/mol K.
These superconducting properties are comparable to those obtained 
with techniques that are not so well suited to industrial scale 
fabrication.

\end{abstract}

\begin{keyword}
\mgb \sep superconductivity
\PACS  74.25.Bt \sep 74.25.Fy \sep 74.25.Ha \sep 74.60.Jg
\end{keyword}
\end{frontmatter}

\section{Introduction}

The discovery of superconductivity in the binary intermetallic
compound \mgb{} with a superconducting critical temperature $T_{c}
\approx 40$ K came as a great surprise and has attracted much
interest.\cite{Nagamatsu01,Bud'ko01,Finnemore01,Canfield01}
This value of $T_{c}$ far exceeds the previous record high
$T_{c}$ value of any binary
intermetallic compound of $23$ K for the A$15$ compound
Nb$_{3}$Ge.\cite{Gavaler73}  Only the oxides
and compounds based on C$_{60}$ have comparable or higher
values of $T_{c}$.\cite{Tanigaki91}  Moreover, this value of
$T_{c}$ is near the limits of $T_{c}$ expected on theoretical grounds
for superconductivity based on the electron-phonon interaction
mechanism.\cite{Varma82,An01}  It is of great interest to study the 
normal- and superconducting-state properties of \mgb{} to gain
insight into the characteristics of the superconducting state
and the underlying electron-pairing mechanism.

Up to date, high quality bulk samples of \mgb{} have been successfully
produced by synthesis under high pressure.
Superconductivity in \mgb{} was discovered by Nagamatsu et al. in 
samples prepared by hot isostatic pressing (HIPing) a mixture of Mg 
and B powders at $196$ MPa.\cite{Nagamatsu01}  
While unequivocally displaying superconductivity, these initial 
\mgb{} samples were quickly surpassed in terms of superconducting 
properties by the compression of commercial \mgb{} powder in a cubic 
multi-anvil-type press with pressures up to $3$ 
GPa.\cite{Jung01,Takano01}
These later results clearly demonstrate that high pressure synthesis is 
advantageous in producing fully dense bulk \mgb{} with 
electrical transport properties and high critical current 
densities superior to those of sintered samples.  However, these
reports also show the limitations on the amount of sample that can
be prepared in each compression cycle.

In this paper, we report electrical transport, magnetization, and
specific heat measurements on high quality bulk \mgb{} samples prepared
from commercial powder by means of a process called dense material
cooling under pressure (DMCUP), a technique based on HIPing that
potentially can be scaled to larger sample sizes and more complex 
shapes.\cite{Indrakanti01}

\section{Experimental Details}

Magnesium diboride (\mgb{}) powders of $-325$ mesh size with $98 \%$ purity 
were obtained from Alfa Aesar, Inc., from which bulk \mgb{} samples
were prepared by HIPing at $200$ MPa as previously reported.\cite{Indrakanti01}  
Magnetization measurements were made with a commercial 
Quantum Design SQUID magnetometer.  DC magnetic susceptibility data 
were taken on a $14$ mg sample in an applied field of $10$ Oe from 
$5$ to $60$ K, using the conventional zero-field-cooled (ZFC) and
field-cooled (FC) procedures. 
Magnetization $M$ vs applied field $H$ hysteresis loops were taken on a
$0.43$ mg sample at temperatures between $10$ and $30$ K and in
applied fields between $-50$ and $50$ kOe.  Electrical resistivity 
measurements were made using the standard four-probe technique in a
commercial Quantum Design Physical Properties Measurement System.
The dc current was applied using a Keithley K220 current source and 
the sample voltage was measured with a Keithley K2182 nanovoltmeter. 

Specific heat measurements were made between $4$ and $90$ K in a
laboratory-built calorimeter using a standard semiadiabatic heat-pulse
technique.  The two samples measured consisted of $(1)$ a $120$ mg
pellet pressed from the original Alfa Aesar powder and $(2)$ three
pieces of the HIPed 
sample with a combined mass of $128$ mg.  The samples were attached to
a sapphire platform using Apiezon N Grease.  The sample
temperature was measured with a Lakeshore Cernox 1030 thermometer and
the heat pulse was applied with a thin-film chip heater; both the
thermometer and the heater were attached to the underside of the
sapphire platform.

\section{Results and Discussion}

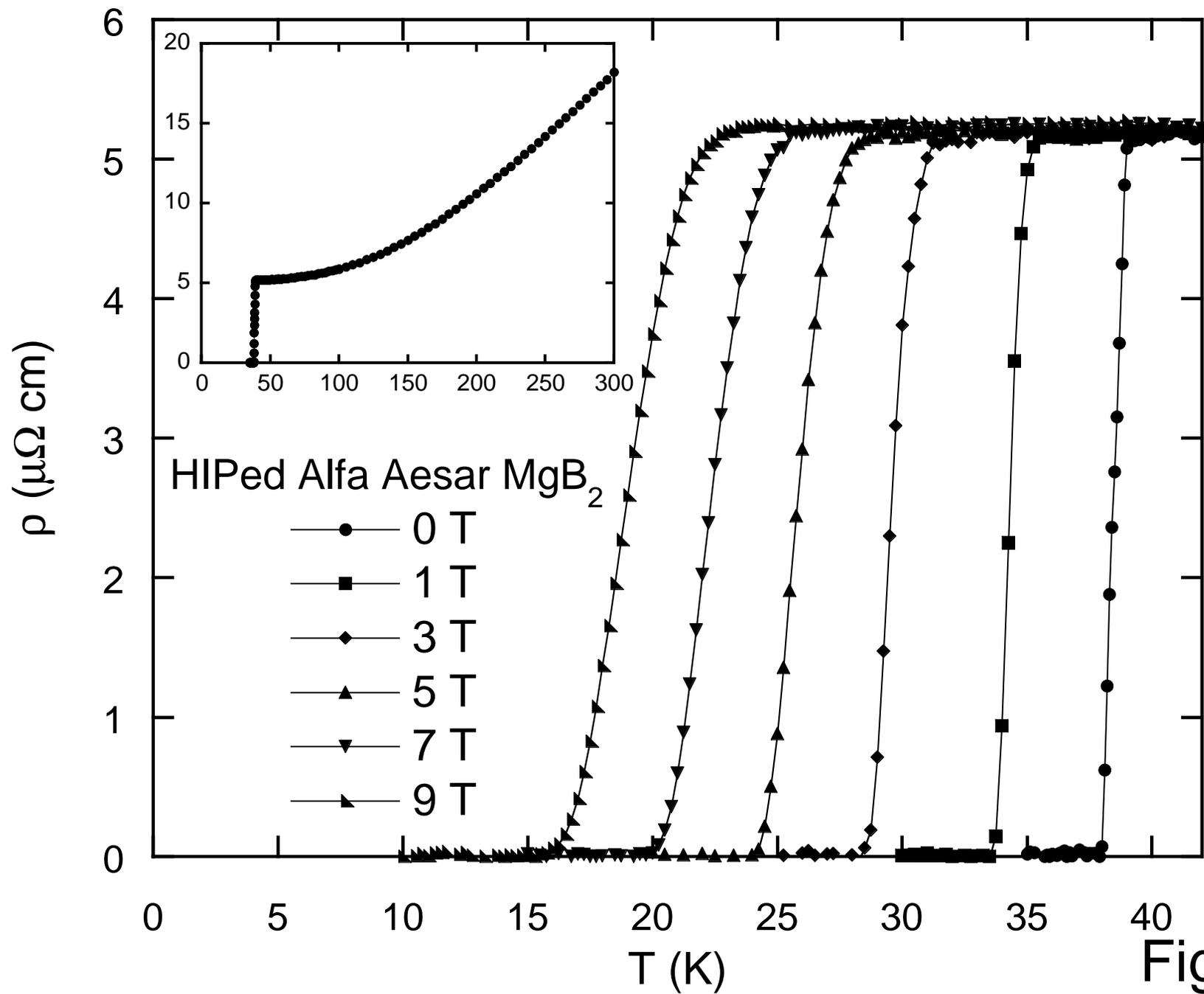
\begin{figure}
    \caption{Electrical resistivity $\rho$ vs temperature $T$ in 
    several magnetic 
fields between $0$ to $9$ T of HIPed Alfa Aesar \mgb.  Inset: 
$\rho$ vs $T$ up to room temperature of HIPed Alfa Aesar 
\mgb.}
\label{R}
\end{figure}

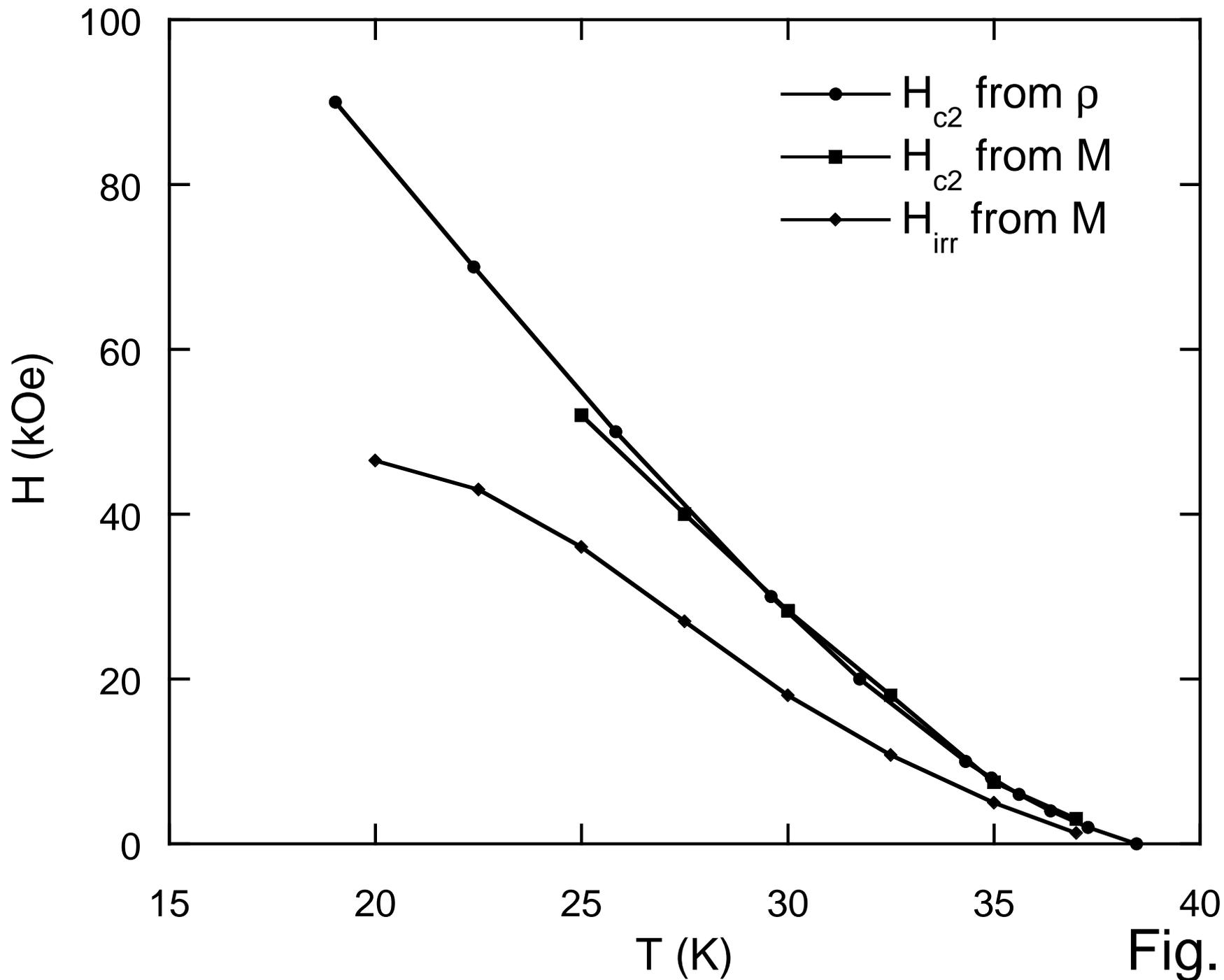
\begin{figure}
    \caption{Critical field $H_{c2}$ and irreversibility field $H_{irr}$
vs temperature $T$ as determined by resistivity and magnetization
measurements of HIPed Alfa Aesar \mgb.}
\label{Hc2}
\end{figure}

The main portion of Figure \ref{R} shows the evolution of the 
superconducting transition of \mgb{} with magnetic field up to $90$ 
kOe as a plot of electrical resistivity $\rho$ vs temperature $T$.  
The resistivity as a function of temperature up to $300$ K is shown in 
the inset to Figure \ref{R}.  At $40$ K, in the normal state right 
above $T_{c}$, the sample has a resistivity of
$5.2$ $\mu\Omega\cdot$cm, which is in the low range of the reported 
resistivity values for \mgb{} prepared from natural boron. The 
$\rho(T)$ data in Figure \ref{R} yield a residual resistivity ratio
[$\equiv{} \rho(300$ K)$/\rho(40$ K)] of $3.46$.
The zero-field superconducting transition occurs at $T_{c} = 38.5$ K,
defined as the temperature of the $50 \%$ value of the transition,
and has a width 
$\Delta{}T_{c} = 0.75$ K, defined as the difference in the 
temperatures of the $10 \%$ and $90 \%$ values of the transition.
The upper critical field $H_{c2}$ vs $T$ curve shown in Figure \ref{Hc2} 
was determined from the $\rho$($T$) curves in the same manner. 
The normal state ($T > T_{c}$) resistivity data display very small
magnetoresistance at high fields, in 
contrast to data previously published on samples synthesized from a 
mixture of high purity Mg and B,\cite{Finnemore01,Fuchs01} but in
agreement with data previously published on samples sintered under
high pressure from commercial $98 - 99 \%$ pure \mgb{}
powder.\cite{Takano01,Dhalle01}  This implies that the
presence of $1 \%$ to $2 \%$ of chemical impurities suppresses the
normal state magnetoresistance almost entirely.

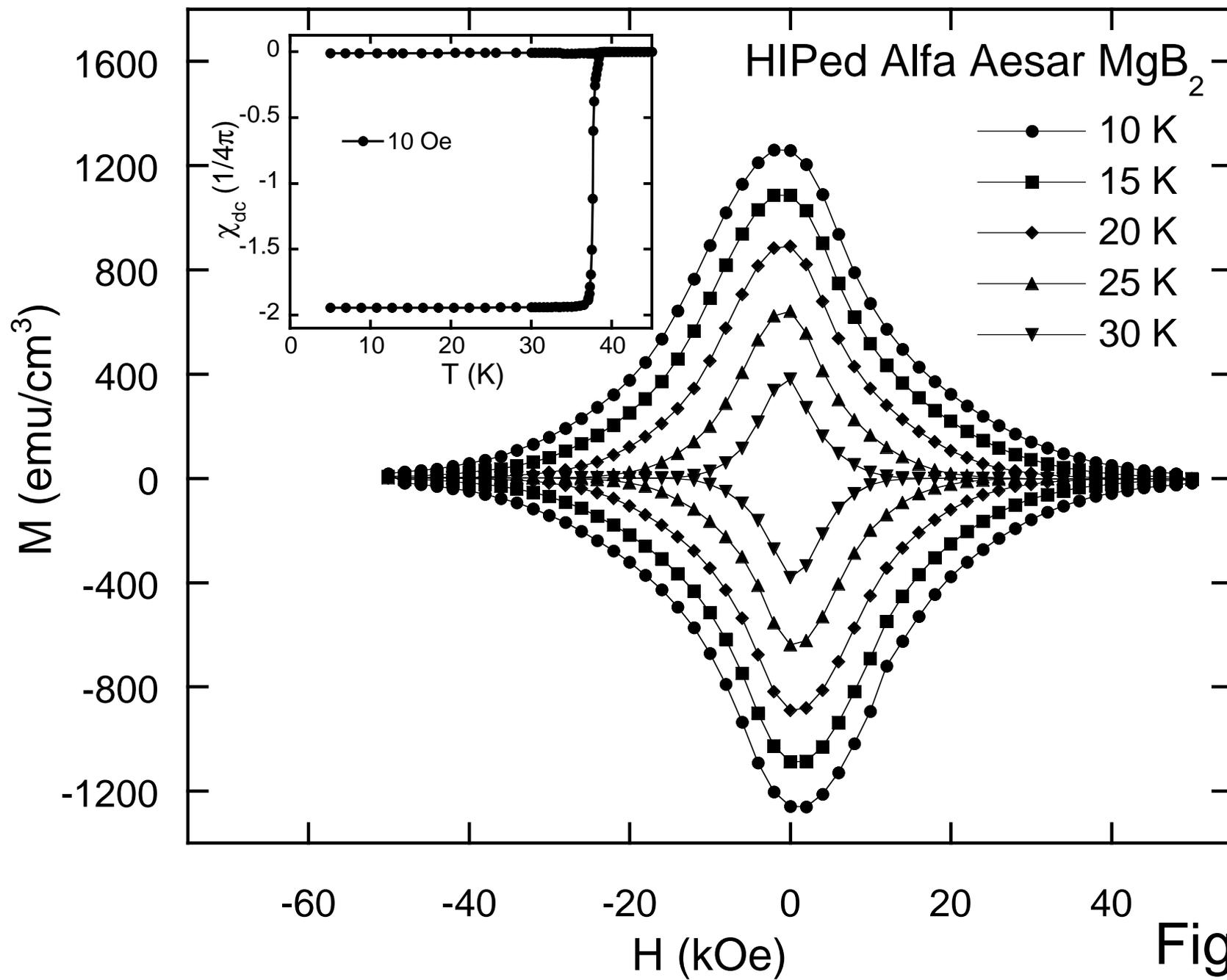
\begin{figure}
    \caption{Magnetization $M$ vs magnetic field $H$ of HIPed Alfa 
Aesar \mgb.  Inset: magnetic susceptibility $\chi_{dc}$ vs 
temperature $T$ of HIPed Alfa Aesar \mgb.}
\label{M}
\end{figure}

Magnetization $M$ vs magnetic field $H$ isotherms for the \mgb{} sample 
are shown in Figure \ref{M}.  Displayed in the inset to Figure \ref{M} is 
a plot of  the dc magnetic susceptibility, $\chi_{dc}$, vs temperature,
$T$, for the \mgb{} sample in a magnetic field of $10$ Oe. 
The very slight tilting of the $M(H)$ data is due to a diamagnetic
background in the normal state which was subtracted to facilitate the 
determination of the $H_{c2}$ data plotted in Figure \ref{Hc2}.  
The high-field magnetization above the superconducting transition 
is described well by a linear function that passes through the origin.
Once these linear functions were subtracted from the $M(H)$ data, the 
critical field $H_{c2}$ could be more conveniently determined as the field
at which the magnetization vanished.  These values of $H_{c2}$ are
extremely close to the values derived from the resistivity measurements,
as shown in Figure \ref{Hc2}.  The irreversibility field, $H_{irr}$,
was easily found from the detailed $M(H)$ measurements as the magnetic
field at which hysteresis sets in.

The critical field at zero temperature can be estimated from the
Werthamer-Helfand-Hohenberg (WHH) theory in the
``dirty limit''\cite{Werthamer66} ($l/\xi \ll 1$, where $l$ is the mean
free path and $\xi$ is the superconducting coherence length) by the formula
$H_{c2}(0) = \frac{1}{2}\sqrt{2}T_{c}|{dH_{c2}/dT}|_{T_{c}}$.  Our 
data as presented in Figure \ref{Hc2} exhibit positive curvature at 
temperatures near $T_{c}$, and becomes linear by $32$ K.  The slope 
from this lower temperature region is $|{dH_{c2}/dT}| = 5.68$ kOe/K;
this yields $H_{c2}(0) = 155$ kOe.  The low 
temperature coherence length can be estimated from the relation
$\xi_{0} = [\phi_{0}/2\pi H_{c2}(0)]^{\frac{1}{2}}$; using the value for 
the flux quantum $\phi_{0}$ of $2.1 \times 10^{-7}$ Oe$\cdot$cm$^{2}$, we 
find $\xi_{0} = 4.6$ nm.  An estimate of the mean free path $l = 4.8$ nm can 
be obtained from the resistivity using the relation 
$l = v_{F}m_{e}/ne^{2}\rho(40$ K), using the following values for the
Fermi velocity $v_{F} \approx 4.8 \times 10^{7}$ cm/s,\cite{Kortus01}
the charge carrier density
$n \approx 6.7 \times 10^{22}$ e/cm$^{3}$,\cite{Canfield01} and the 
resistivity $\rho$ ($40$ K) $\approx 5.2$ $\mu\Omega\cdot$cm.  The 
resulting ratio $l/\xi \approx 1$ implies that the ``dirty limit'' is 
not really appropriate in the \mgb{} specimen studied.  However, it should be
pointed out that many of the published results that are well into
the ``clean limit'' ($l/\xi \gg 1$) continue to use the ``dirty limit''
approximation to estimate $H_{c2}(0)$.

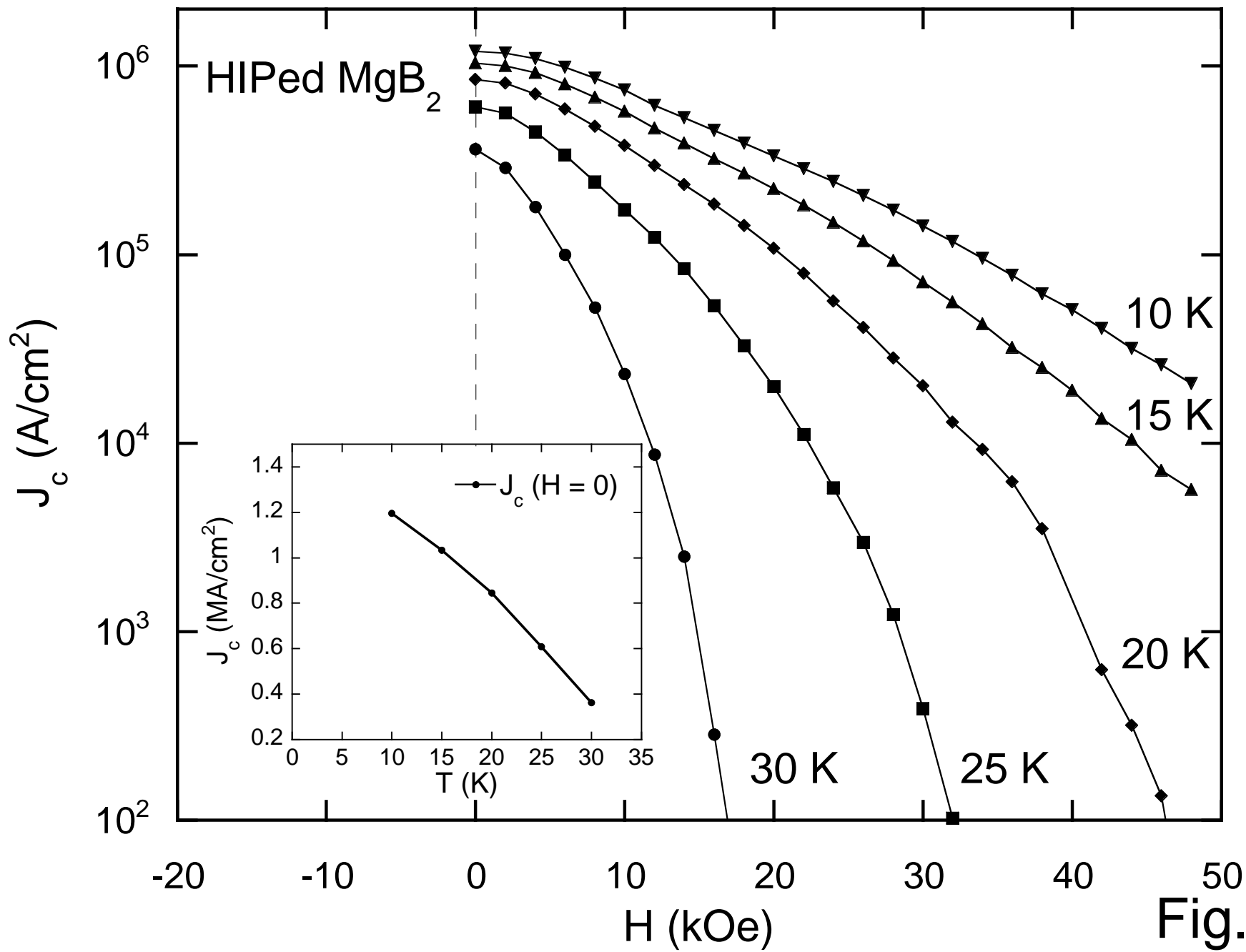
\begin{figure}
    \caption{Critical current density $J_{c}$ as a function of magnetic 
field $H$ of HIPed Alfa Aesar \mgb.  Inset: critical current density 
$J_{c}$ as a function of temperature $T$ of HIPed Alfa Aesar \mgb.}
\label{Jc}
\end{figure}

The critical current density $J_{c}$ of the \mgb{} sample was determined
from the magnetization data by applying the Bean critical state
model,\cite{Bean64} according to which 
$J_{c}$ is linearly related to the magnetization $M$ by a 
proportionality constant which depends only on the dimensions of the 
sample being measured.  Our triangular sample required the formula
$J_{c} = 15s\Delta{}M/A$; where $s = \frac{1}{2}(a+b+c)$
is the semiperimeter of the triangle; $\Delta{}M$ is the width of the
magnetization loop; and $A$ is the area of the triangle.\cite{Ji92}
The results from applying this model are shown in the main portion of 
Figure \ref{Jc}.  The value of the critical current density at low
temperatures is extremely high, almost an order of magnitude greater
than previously reported.\cite{Finnemore01,Takano01,Dhalle01}  
The inset to Figure \ref{Jc} shows a plot of the critical current density at
zero applied field versus temperature.  The concavity of the curve
seems to indicate a zero-temperature value for $J_{c}$ of $\sim 1.4$
MA/cm$^{2}$.  The rate at which $J_{c}$ decreases with increasing field
is also much slower compared to previously reported results.

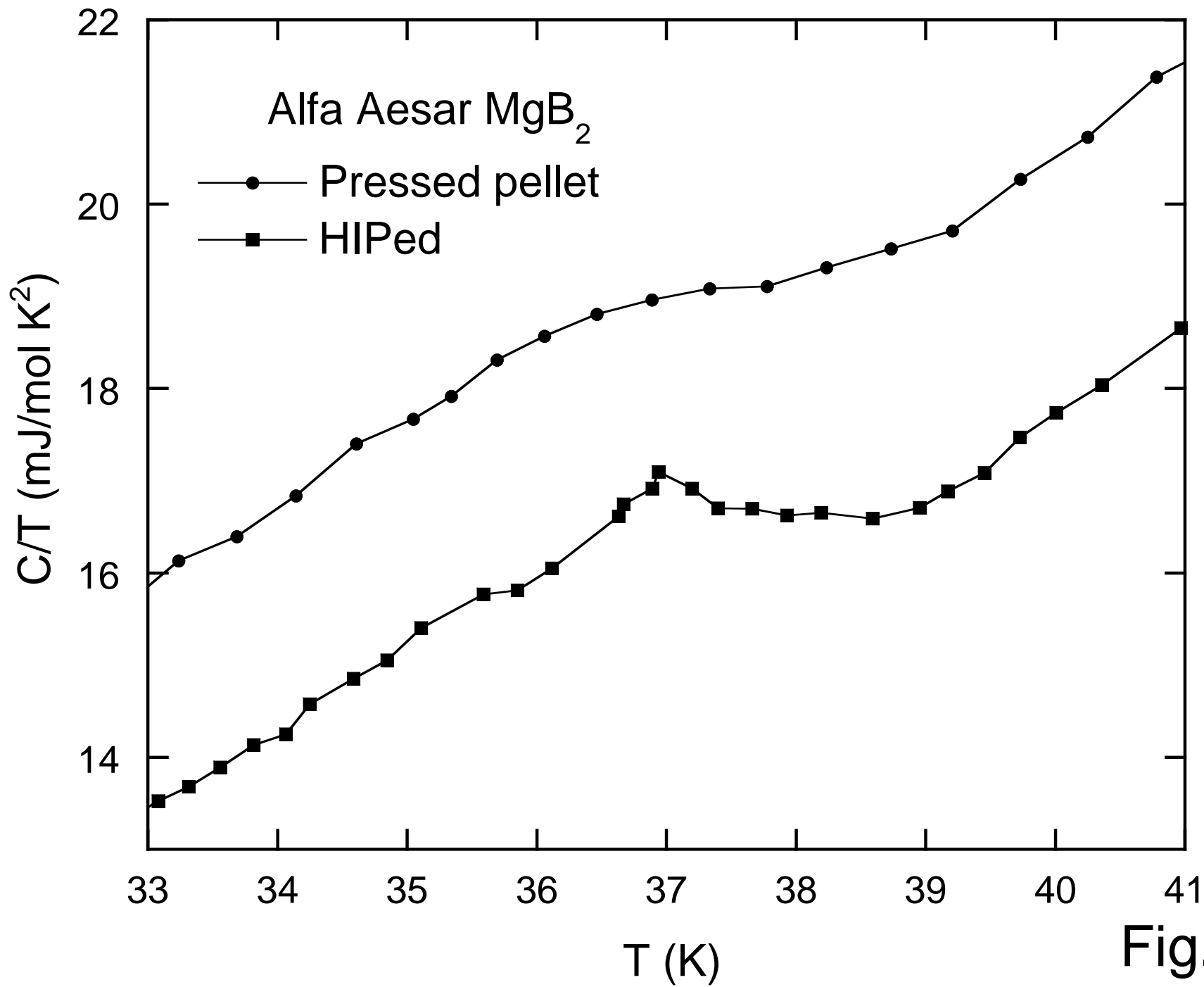
\begin{figure}
    \caption{Specific heat divided by temperature $C_{p}/T$ vs $T$ of 
    powder and HIPed samples of Alfa Aesar \mgb.}
\label{Cp}
\end{figure}

Figure \ref{Cp} compares the specific heat divided by temperature 
$C_{p}/T$ vs $T$ curves for both a pressed pellet of as-received
Alfa Aesar \mgb{} powder and HIPed \mgb{}.  The HIPed sample exhibits
a sharper superconducting transition as well as a smaller specific heat. 
The data for the HIPed sample yielded a superconducting transition
temperature $T_{c}$ of $38.1$ K and a specific heat jump
$\Delta{}C$ of $92$ mJ/mol K.  These values, as well as the 
shape of the curve, are comparable to previously published specific 
heat data.\cite{Bud'ko01,Kremer01}  Since electrical transport and 
magnetization data were greatly improved by preparing \mgb{} at high 
pressures, it appears that polycrystalline \mgb{} simply does not
display a sharp superconducting transition in its specific heat.
At the present time we are unable to
suppress the superconducting transition with a magnetic field in our
calorimeter, and so a full comparison with previously published
specific heat data\cite{Bud'ko01,Kremer01} will have to await a 
future publication.
However, it should be noted that with the published Debye temperature
$\Theta_{D}$ of $\sim800$ K, a true fit of the equation $C = \gamma T
+ \beta T^{3}$ may not be valid above $\Theta_{D}/50 \approx 16$ K.
This would require specific heat measurements in magnetic fields well
above $9$ T, which have not yet been carried out.

\section{Summary}
In summary, we have studied the superconducting properties of fully
dense \mgb{} bulk specimens prepared using a procedure called DMCUP
that is based on HIPing.
Our samples displayed very small magnetoresistance in high fields, 
consistent with other measurements on samples prepared from 
commercial powder, as well as a narrow superconducting transition 
width $\Delta{}T_{c}$.  We also observed large values of the upper 
critical field $H_{c2}$ and the critical current density $J_{c}$, and 
estimated these properties at $T = 0$ K.  A comparison of the specific 
heat between powdered and HIPed \mgb{} samples showed only a small 
improvement in the specific heat jump $\Delta{}C$, suggesting that 
it may only be possible to observe a sharp specific heat jump for 
single crystal specimens of \mgb.  The HIPing
technique used to prepare the \mgb{} sample has great potential as
it is easily scaled up for larger samples.

We would like to thank E. D. Bauer, E. J. Freeman, and W. M. Yuhasz 
for valuable discussions.  This research was supported by the U.S.
Department of Energy under Grant No. DE FG03-86ER-45230.


\pagebreak

\end{document}